\documentclass[aps, prb, twocolumn, superscriptaddress, english]{revtex4-2}

\usepackage{graphicx}
\usepackage{cancel}
\usepackage{enumerate}
\usepackage{hyperref}
\usepackage{textcomp}
\usepackage{amsmath}
\usepackage{amssymb}
\usepackage{pgf}
\usepackage{soul}
\usepackage{subfigure}
\usepackage[english]{babel}
\usepackage[normalem]{ulem}

\def\ket#1{| #1\rangle}

\def\ud{\mathrm{d}}

\newcommand{\nep}{\textrm{e}}

\newcommand{\opc}[1]{{\hat{c}^{\phantom \dagger}}_{#1}}
\newcommand{\opcdag}[1]{{\hat{c}^{\dagger}}_{#1}}
\newcommand{\opd}[1]{{\hat{d}^{\phantom \dagger}}_{#1}}

\newcommand{\opn}[1]{\hat{n}_{#1}}

\newcommand{\Ham}{\widehat{H}}

\newcommand{\Up}{{\uparrow}}
\newcommand{\Dn}{{\downarrow}}

\graphicspath{{./}{./Figure/}}

\begin{document}
\title{Simulations of the dynamics of quantum impurity problems with matrix product states}
\author{Matteo M. Wauters}
\affiliation{Niels Bohr International Academy and Center for Quantum Devices, Niels Bohr Institute, Copenhagen University, Universitetsparken 5, 2100 Copenhagen, Denmark}
\affiliation{INO-CNR BEC Center and Dipartimento di Fisica, Università di Trento, 38123 Povo, Italy}
\affiliation{INFN-TIFPA, Trento Institute for Fundamental Physics and Applications, Trento, Italy}

\author{Chia-Min Chung}
\affiliation{Department of Physics, National Sun Yat-sen University, Kaohsiung 80424, Taiwan}
\affiliation{Center for Theoretical and Computational Physics, National Sun Yat-Sen University, Kaohsiung 80424, Taiwan}
\affiliation{Physics Division, National Center for Theoretical Sciences, Taipei 10617, Taiwan}

\author{Lorenzo Maffi}
\affiliation{Niels Bohr International Academy and Center for Quantum Devices, Niels Bohr Institute, Copenhagen University, Universitetsparken 5, 2100 Copenhagen, Denmark}

\author{Michele Burrello}
\affiliation{Niels Bohr International Academy and Center for Quantum Devices, Niels Bohr Institute, Copenhagen University, Universitetsparken 5, 2100 Copenhagen, Denmark}

\begin{abstract}

The Anderson impurity model is a paradigmatic example in the study of strongly correlated quantum systems and describes an interacting quantum dot coupled to electronic leads. Here we investigate its dynamics following a quantum quench based on matrix product state simulations. We examine the behavior of its impurity magnetization. Its relaxation allows us to extract the predicted scaling of the Kondo temperature as a function of the impurity-lead hybridization and quantum dot repulsion. Additionally, our simulations provide estimates of the currents in the nonequilibrium quasi-steady state appearing after the quench. Through their values, we examine the dependence of the conductance on the voltage bias $V_b$ and on the impurity chemical potential $V_g$, which displays a zero-bias Kondo peak. Our results are relevant for transport measurements in Coulomb blockaded devices, and, in particular, in quantum dots induced in nanowires.
\end{abstract}

\maketitle

\section{introduction}\label{sec:intro}

The Kondo effect is the most emblematic embodiment of strong correlations in condensed matter systems. The advances in the fabrication and measurement techniques of nanostructures allowed us to observe its distinctive zero-bias conductance peak in a wide class of systems, including gate-defined quantum dots \cite{Goldhaber1998,Cronenwett1998}, nanotubes \cite{Nygaard2000} and semiconducting nanowires \cite{Jespersen2006}.

In these mesoscopic systems, however, the dynamics of the quantum impurities at the basis of the Kondo effect is typically too fast to be observed.
A complementary experimental platform has been recently offered by quantum simulators of ultracold fermionic Yb atoms \cite{Riegger2018}. In these setups, the characteristic time scales are much longer than in their solid state counterpart, thus enabling the analysis of the dynamics of the spin impurities at the basis of the Kondo effect in out-of-equilibrium transient states \cite{Demler2018}.

Inspired by these developments, in this work we analyze the dynamics of the Anderson impurity model after a quantum quench through matrix product state (MPS) simulations. By studying the transient behavior of its impurity magnetization, we provide a numerical verification of the Kondo time scale consistent with previous renormalization group results \cite{Haldane1978}. We derive 
the conductance of the corresponding two-terminal problem by considering both leads with a finite voltage bias and different charge impurity configurations. Our results are relevant for the experimental study of Coulomb blockaded nanowires with induced quantum dots.

The out-of-equilibrium properties of quantum impurity models following quantum quenches are considered a paradigmatic playground to observe how strong correlations develop through time evolution in many-body quantum systems and have been recently studied by means of a vast set of analytical and numerical techniques. The quench dynamics of the two-terminal Anderson impurity model (AIM), in particular, has been addressed by quantum Monte Carlo methods to estimate the related Keldysh Green's functions \cite{Millis2009,Werner2010,Gull2011,Profumo2015,Gull2019,Gull2022}, time-dependent numerical renormalization group \cite{Anders2008}, self-consistent diagrammatic techniques \cite{Souto_NJP2018}, and tensor-network methods to study its real-time evolution \cite{Dagotto2006,DaSilva_PRB2008,PhysRevB.79.235336,Weichselbaum2018,Manaparambil_PRB2022,Kohn_2022,Thoenniss2023} (see also the comparison among different methods in Ref. \cite{Eckel2010}).

Despite being one of the most studied interacting models in condensed matter physics, the AIM keeps drawing attention as a standard test-bed for both numerical and analytical investigations of out-of-equilibrium strongly correlated systems and non-perturbative effects.

In the following we will benchmark the MPS algorithm we introduced in Ref. \cite{Chung_PRB2022} through its simulation of the time evolution of the two-terminal Anderson impurity model (AIM). Our approach is tailored to examine the Kondo regime and make predictions on its transport properties in the low-temperature limit. Additionally, it can be readily extended to encompass more intricate forms of hybridization between the leads and the impurity and characterize transport features through more complex structures than the AIM, such as multilevel or superconducting quantum dots and different kinds of interacting scatterers.

The rest of the paper is organized as follows. We introduce the model in Sec.~\ref{sec:model} and we give and overview of our numerical approach in Sec.~\ref{sec:MPS}, summarizing the main concepts of Ref.~\cite{Chung_PRB2022}.
We illustrate our results in Sec.~\ref{sec:rsults}, where we focus on the estimate of the Kondo temperature from the relaxation of the magnetization and on the current-voltage curves. 
We summarize our work and draw our conclusions in Sec.~\ref{sec:conclusion}.
In the appendix~\ref{app1} we make an explicit comparison between the spin relaxation computed with our approach and the results presented in Ref.~\cite{Anders2005} using the time-dependent numerical renormalization group algorithm.

\section{The model}\label{sec:model}
The AIM represents an electronic environment coupled with an interacting magnetic impurity; it is one of the most popular yet simple models that display the Kondo effect, and it constitutes the central element of dynamical mean field theory methods for studying correlated materials, making it a fundamental problem for many numerical algorithms~\cite{Aoki_RevModPhys2014}.
\begin{figure}
    \centering
    \includegraphics[width=\columnwidth]{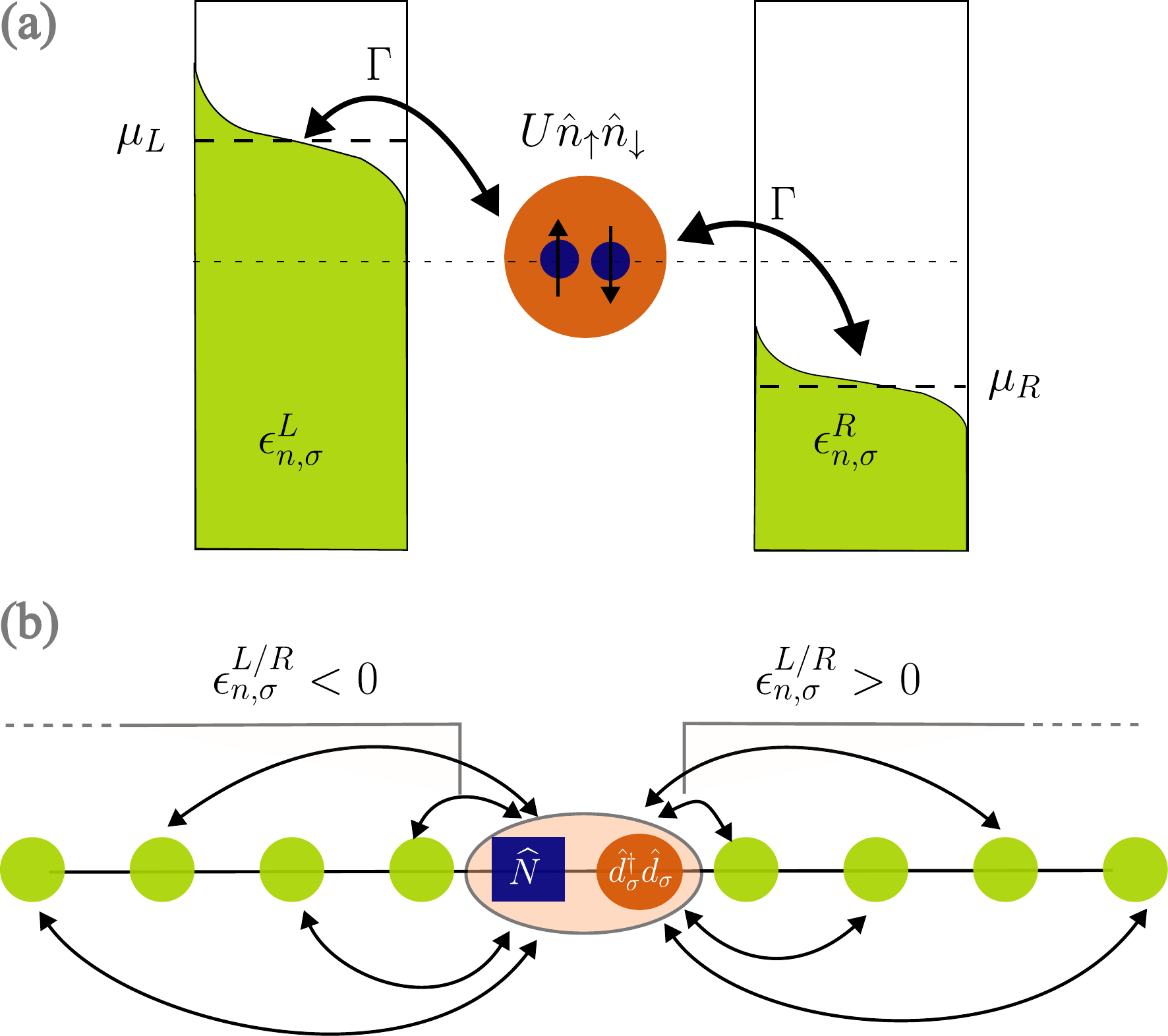}
    \caption{(a) Sketch of the AIM: a single-level quantum dot with Hubbard repulsion $U$ is tunnel-coupled to two noninteracting leads with chemical potentials $\mu_L$ and $\mu_R$.
    (b) Schematic representation of the MPS describing the system \cite{Chung_PRB2022}. The sites of the chain represent single-particle orbitals and are ordered by their energy. To account for the interaction, we include an auxiliary bosonic charge site (represented by a square) which counts the number of particles inside the dot. This construction introduces long-range couplings (arrows) in the Hamiltonian MPO, which however do not constitute an obstacle for the TDVP algorithm used for the time evolution.}
    \label{fig:AIM_sketch}
\end{figure}
Its Hamiltonian reads
\begin{equation}
    \Ham = \Ham_{\rm leads} + \Ham_{\rm tunn} + \Ham_{\rm AIM}\ ,  \label{Htot}
\end{equation}
where 
\begin{equation}
    \Ham_{\rm AIM} = U \opn{\uparrow}\opn{\downarrow} + V_g (\opn{\uparrow}+\opn{\downarrow})\,, 
\end{equation}
with $\opn{\sigma}=\hat{d}_\sigma^\dag \opd{\sigma}$ describing the occupation of the two spin states of  a single-level quantum dot, which, in turn, plays the role of the magnetic impurity and is characterized by the Hubbard repulsion $U$ and the chemical potential $V_g$.
Unless otherwise stated, we will focus on the particle-hole symmetric point $V_g = -0.5 U$.

The lead Hamiltonian
\begin{equation}
    \Ham_{\rm leads}= -\sum_{\alpha, \sigma,  l} t_{\alpha,\sigma,l} \left( \opcdag{\alpha,\sigma,l}\opc{\alpha,\sigma,l} +{\rm H.c.} \right)+ \sum_{\alpha, \sigma,  l}\mu_{\alpha,\sigma}\opn{\alpha,\sigma,l}
\end{equation}
describes two spinful fermionic chains ($\alpha=L,R$) with a spin-dependent chemical potential $\mu_{\alpha,\sigma}$ and a hopping amplitude $t_l = t_0 \nep^{-(l-1)/\xi}$ that decreases exponentially as a function of the distance from the site $l=1$, with a decay length $\xi$. This is known as Wilson construction and it is commonly used in numerical renormalization group approaches to impurity problems. Moreover, it has been shown effective to increase both the resolution at small voltage bias, namely by mimicking an effectively larger system, and the stability of the time evolution in MPS simulation of transport problems  \cite{DaSilva_PRB2008,Chung_PRB2022,Weichselbaum2009}. Indeed, given the finite size $L$, the density of states around the Fermi energy depends on the hopping decay length $\xi$: the smaller $\xi$, the more states are shifted toward the Fermi energy, leading to a smaller energy level spacing. Therefore, a strong decay of the tunneling provides higher energy resolution to accurately determine the dynamics for states close to zero energy (thus at small bias voltages)\cite{DaSilva_PRB2008}. 

Finally, the quantum dot and the leads are coupled with a standard tunneling Hamiltonian
\begin{equation}
    \Ham_{\rm tunn} = -\sum_{\alpha,\sigma} J_{\alpha,\sigma} \left( \opcdag{\alpha,\sigma,1}\opd{\sigma}+{\rm H.c.}\right) \ ,
\end{equation}
where $\opd{\sigma}$ destroys an electron with spin $\sigma$ on the impurity level. Throughout this paper, we consider a uniform tunneling strength between the quantum dot and the leads $J_{\alpha,\sigma} = J$ and we denote by $\Gamma=2J^2/t_0$ the effective tunneling rate in the limit of infinite bandwidth (constant density of states).

To bring the system out of equilibrium, we adopt two different quantum quench protocols \cite{Chung_PRB2022,Chien2014}: 
{\em (i)} in the {\em zero-bias quench}, we initialize the system with $J=0$ and $\mu_L = \mu_R$, thus preparing a product state between the impurity and the leads; at time ${\sf t}>0$, the leads are connected to the quantum dot ($J>0$) and the system equilibrates towards a stationary state.
{\em (ii)} in the $\mu-${\em quench}, the system is initialized in the ground state at half filling, i.e. with uniform chemical potential $\mu_L=\mu_R$, and then it evolves in time after a voltage bias $V_b$ is turned on.
The first protocol is more useful to study the relaxation of the impurity magnetization and extract the Kondo temperature, while the second leads to a fast convergence of the current to a nonequilibrium quasi-steady state (NEQSS) that describes the dynamics of the quantum quenches of finite-size systems at intermediate times.
The time-evolution of the system after a $\mu-${quench} is indeed analogous to the dynamics of quantum quenches describing the melting of domain walls in spin systems \cite{bertini2016,Essler_2016} and is characterized by the onset of a light-cone departing from the quantum dot at ${\sf t}=0$. The corresponding post-quench Hamiltonian obeys a Lieb-Robinson bound \cite{liebrobinson,bonnes2014} such that, for ${t >0}$, the central region of the system (in proximity of the dot) is not affected by finite-size effects until the quasi-particles produced by the quench do not reach the boundary $l=L$ of the finite leads. As verified in different models with fermionic leads, this implies that the conductance of the system can be extracted by the currents in the NEQSS \cite{Viti_2016,Ljubotina_SciPost2019,Chung_PRB2022}. 
Furthermore, since in the $\mu$-quench the initial state already captures some of the non-perturbative Kondo correlations, such state is closer to the Kondo-like quasi-steady state that arises in transport measurements and it provides the most efficient approach to capture the behavior of the currents in the transient NEQSSs.

\section{Matrix product state simulations}\label{sec:MPS}
Tensor networks offer a powerful framework to simulate the real-time evolution of quantum impurity models \cite{Dagotto2006,DaSilva_PRB2008,Schmitteckert2008,PhysRevB.79.235336,Guttge2013,He_PRB2017,Weichselbaum2018,Thoenniss2023b,Thoenniss2023,Manaparambil_PRB2022,Kohn_2022,Ng2023}. 
The MPS construction enhances the characterization of spectral features at high frequencies, which hold significant importance in out-of-equilibrium and time-dependent impurity problems \cite{Weichselbaum2009}. 

In our simulations, we will describe the time-evolution of the closed and finite system defined by the Hamiltonian \eqref{Htot} with leads of size $L$. As mentioned above, we will focus our analysis to intermediate times after the quantum quench, such that our results will be derived from NEQSSs which are insensitive to finite-size effects. This implies, for instance, that, differently from time-dependent NRG procedures, our MPS results are unaffected by the incomplete thermalization caused by the discretized Wilson chain discussed in Ref. \cite{Rosch2012} because they are unrelated to the thermal Gibbs state emerging at infinite time.

Finite-size effects, however, set important limitations that we must consider. First, in order to avoid boundary effects, our simulations must be restricted to a finite time window set by the upper bound ${\sf t}_{\rm max}$ when the quasiparticles produced during the quantum quench reach the edges of the boundary and, for instance, cause a non-physical reflection of the current after a $\mu$-quench. This problem is mitigated by the adopted Wilson chain approach, which increases the typically accessible time-scale before observing edge effects, ${\sf t}_{\rm max}= \frac{\xi}{2t_0}\left(e^{\left(L-1\right)/\xi}-1\right)$ \cite{Chung_PRB2022}. In the uniform chain case instead, the current travels with constant speed $v_f=2t_0$ through the leads, leading to an edge reflection after the time interval $\left(L-1\right)/2t_0$. 
Additionally, finite-size effects determine also an incomplete relaxation of the spin of the impurity (see next section) after a zero-bias quench, which, again, sets a limit to the physical time window we can consider.
Finally, they also determine the resolution we can achieve in the chemical potentials of the leads and, therefore, in the introduction of a voltage bias in the system to estimate its conductance.

To simulate the post-quench dynamics, we model the system with the MPS depicted in Fig. \ref{fig:AIM_sketch}(b): each site represents a single-particle {\em energy} orbital of the non-interacting and decoupled system ($U, J =0$), and we compute the unitary time evolution of the closed system with the time-dependent variational principle (TDVP) \cite{Haegeman_PRL2011,Haegeman2016}.

In particular, we expand the construction presented in Ref.~\cite{Chung_PRB2022} with the addition of the spin degrees of freedom; the MPS "sites" are ordered based on their energies \cite{Rams2020}, such that the entropy growth during the time evolution is restricted in an energy window, thus a segment of the MPS, corresponding to the voltage bias. 
This can be intuitively understood by considering that in transport processes, only states in an energy window $[\epsilon_F-eV_b, \epsilon_F+eV_b]$, where $\epsilon_F$ is the Fermi energy of the leads, contribute to the current. All the others remain in their initial state, occupied or empty, and do not contribute to the entanglement growth in the MPS. In particular, as pointed out in Ref. \cite{Rams2020}, the entanglement entropy displays a logarithmic growth with time instead of a linear increase.
Since the basis states (MPS "site") are ordered by their energies regardless the number of leads, introducing multiple leads (or the spin degrees of freedom) is straightforward.

The interaction $U$ is introduced  by including an auxiliary MPS site that represents the charge $\widehat{N}=\hat{n}_\Up + \hat{n}_\Dn$ of the dot \cite{Chung_PRB2022}. Tunneling events increase or decrease this charge by one. This construction is not strictly necessary for a single impurity site as in the AIM in Eq. \eqref{Htot}, but it can easily allow for generalizations to multilevel dots with a uniform all-to-all Coulomb repulsion described by an effective charging energy. Moreover, representing separately the charge degree of freedom from the dot particle number enables us to consider superconducting quantum dots\cite{Chung_PRB2022,Wauters_PRB2023}.

In the chosen single-particle energy basis, $\Ham_{\rm leads} + \Ham_{\rm AIM}(U=0)$ acts trivially on each site of the MPS as they represent eigenstates of the Hamiltonian with $U,\ J =0$. Therefore the dynamics is dictated only by the interplay between the tunneling Hamiltonian coupling the leads with the quantum dot and the Hubbard repulsion $U\opn{\uparrow}\opn{\downarrow}$. $\Ham_{\rm tunn}$ is nonlocal in this basis, but it can be described by a matrix product operator (MPO) with limited bond dimension $\chi=8$, such that TDVP is not hampered by the presence of these long-range tunneling elements and can be efficiently used to simulate the dynamics for a long evolution time.

The method is implemented by using ITensor library~\cite{itensor}.
The source code can be found in Ref. \footnote{The source code can be found in the repository \url{https://github.com/chiamin/QuenchTranportTwoChains}}.

\section{Results}\label{sec:rsults}
We first focus on the equilibration of the impurity after it is coupled to the {\em unbiased} leads (zero-bias quench). 
In strongly correlated quantum impurity models, the dynamics of the impurity magnetization is typically characterized by two rates: $\Gamma$, which determines the short-time and nonuniversal evolution; and the Kondo temperature $T_K$, whose inverse, the Kondo time $\mathsf{t}_K = T_K^{-1}$, defines the time scale required for the formation of the Kondo screening cloud (see, for instance, Ref. \cite{Anders2005} for the Anderson impurity model coupled with a single lead, and Ref. \cite{Lobaskin2005} for the corresponding Kondo problem). 
In the renormalization group sense, the evolution for time $\Gamma^{-1}<\mathsf{t}\lesssim\mathsf{t}_K$ is governed by the weak-coupling fixed point of the Kondo problem \cite{Cavalcante_PRB2023}, and $\mathsf{t}_K$ constitutes the decay time of the magnetization in this intermediate regime towards the formation of a spin singlet with the conduction electrons.

Therefore, we aim to get an estimate of the Kondo temperature as a function of the ratio between the interaction strength $U$ and the effective tunneling rate $\Gamma$ from the dynamics of the impurity magnetization $\langle \sigma^z \rangle=\langle \opn{\uparrow}\rangle -\langle \opn{\downarrow}\rangle  $.
We prepare the quantum dot in the polarized state $|\opn{\Up}=1,\opn{\Dn}=0\rangle $ and measure its evolution in time after a zero-bias quench.
For this analysis, we choose $L=64$ as the lead length and the hopping decay length between $\xi=8$ and $\xi=32$, depending on the energy resolution needed to accurately measure the magnetization up to times of the order of ${\sf t}_K$. 

We consider two values for the interaction strength, $U=t_0$ and $U=0.4t_0$, and we examine the particle-hole symmetric point $V_g=-0.5U$. To extract the predicted exponential dependence of the Kondo temperature from $U/\Gamma$ \cite{Wang_PRB2008,He_PRB2017, Cavalcante_PRB2023}, we vary the hybridization strength $\Gamma$ between $\sim U/20$ ($J\sim 0.15 U$) and $\sim U/2$ ($J\sim 0.5U$).

\begin{figure}
    \centering
    \includegraphics[width=\columnwidth]{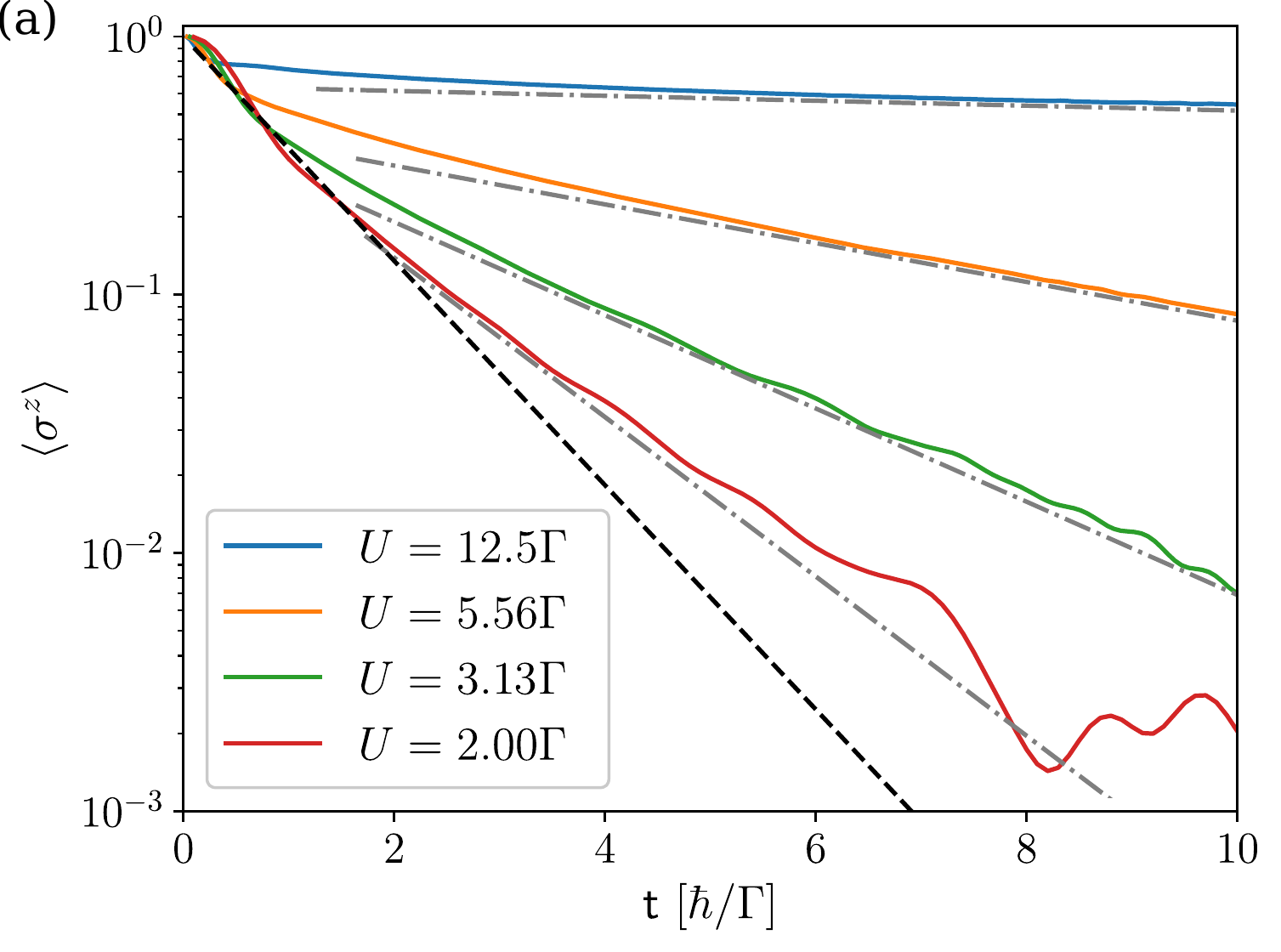}
    \includegraphics[width=\columnwidth]{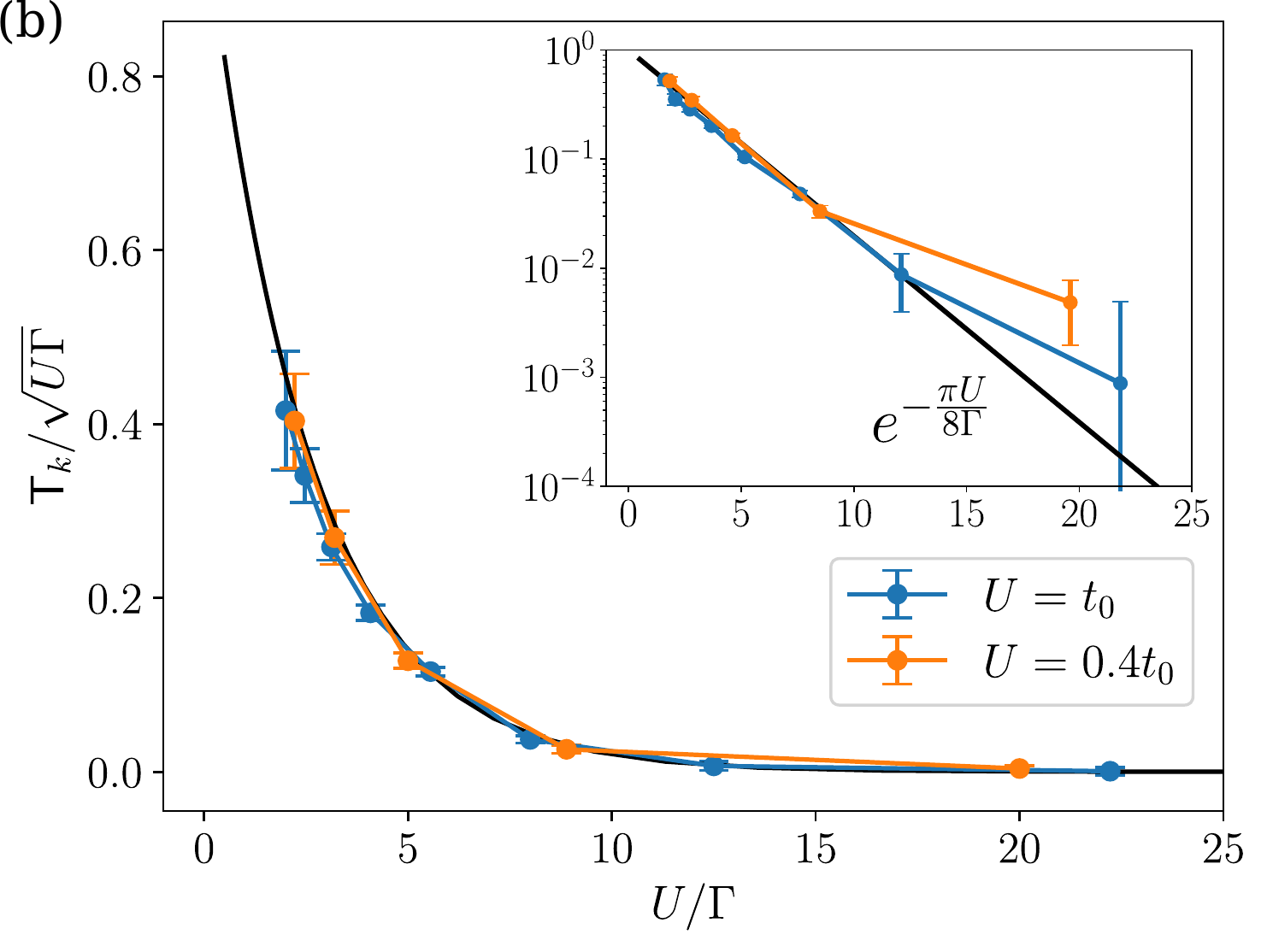}
    \caption{(a): Expectation value of the magnetization $\langle \sigma^z\rangle  = \langle\opn{\uparrow} \rangle-\langle\opn{\downarrow} \rangle$ as a function of time (in units of $\hbar/\Gamma$) in a zero-bias quench. The dashed black lines indicates the short-time relaxation $\nep^{-{\sf t}\Gamma/\hbar}$ while the gray dot-dashed lines highlight the slow dynamics due to the Kondo resonance $\nep^{-{\sf t}/{\sf t}_K}$. The data correspond to $U=t_0$
    (b): Kondo temperature, extracted from the slow relaxation shown in panel (a), versus the effective interaction strength for two values of $U$. The solid black line corresponds to the RG prediction for the Kondo temperature $T_K \sim \sqrt{U\Gamma} \nep^{-\frac{\pi U}{8\Gamma}}$.
    The inset shows the same data in logarithmic scale to emphasize the exponential behaviour of $T_K$.
    For large values of $U/\Gamma$ our accuracy is limited by the small signal-to-noise ratio of the time evolution of the magnetization.}
    \label{fig:kondo_relax}
\end{figure}

Figure \ref{fig:kondo_relax}(a) shows the decay in time of the magnetization for different values of $U/\Gamma$ while we fix $U=t_0$. We can easily identify three regimes: at short time ${\sf t}\lesssim \Gamma^{-1}$, the different curves collapse on each other as the relevant time scale for the relaxation of the impurity is set only by $\Gamma$ (dashed black line). Indeed, notice that time is measured in units of ${\Gamma^{-1}}$.
At longer times, the relaxation rate depends on the ratio $U/\Gamma$, with a slower decay the further the system lies in the strongly interacting/weak-coupling regime.
For these intermediate values of $\mathsf{t}$, we can extract the relaxation time by exponential fits of the data (dot-dashed gray lines).
Finally, the impurity approaches a steady state with a finite magnetization; in Fig.~\ref{fig:kondo_relax}(a) this last regime is visible only for $U=2\Gamma$. Due to the unitary dynamics, the system keeps the memory of its initial state and a complete relaxation to a $SU(2)$ invariant state can not be reached. Indeed, the initial state is prepared with unpolarized leads and the impurity in the $\ket{\uparrow}$ state, leading to a total magnetization of 1. as the system relaxes, we expect a residual impurity magnetization $\langle \sigma^z \rangle \sim (2L)^{-1}$, assuming it spreads over the whole system.
Comparable results have been obtained for the AIM with a single lead in Ref. \cite{He_PRB2017} through the real-time density matrix renormalization group (DMRG) applied to a superposition of four MPS which results in a logarithmic growth of the entanglement similarly to our energy-basis construction.

Figure \ref{fig:kondo_relax}(b) illustrates the inverse of the relaxation times $\mathsf{t}_K(U/\Gamma)$ extracted from the magnetization decay at intermediate times [grey lines of panel (a)] as a function of $U/\Gamma$ and for two values of the Hubbard  interaction $U$. We interpret this quantity as the Kondo temperature $T_K \sim \mathsf{t}_K^{-1}$. 
A comparison with the renormalization group prediction 
\begin{equation}
T_K \sim \sqrt{U\Gamma} \nep^{-\frac{\pi U}{8\Gamma}}
\end{equation}
(solid black line) shows excellent agreement with our data for both values of the interaction strength. 
In the inset, the same data are displayed in logarithmic scale to emphasize the exponential dependence of the Kondo temperature from $U/\Gamma$.  Moreover, the two datasets perfectly collapse on top of each other, highlighting the universal character of the exponential decay linked to the Kondo temperature. 
At very weak coupling ($U/\Gamma \gg 1$) the long evolution time needed for an accurate estimate of the relaxation time can not be reached and our data deviate from the analytical prediction.
Although the entanglement growth ultimately limits our ability to simulate the evolution of large systems at long time, thus preventing the observation of Kondo correlations for very weak coupling, our method allows us to observe nonperturbative effects emerging directly from the nonequilibrium properties of the AIM.
\begin{figure}
    \centering
    \includegraphics[width=\columnwidth]{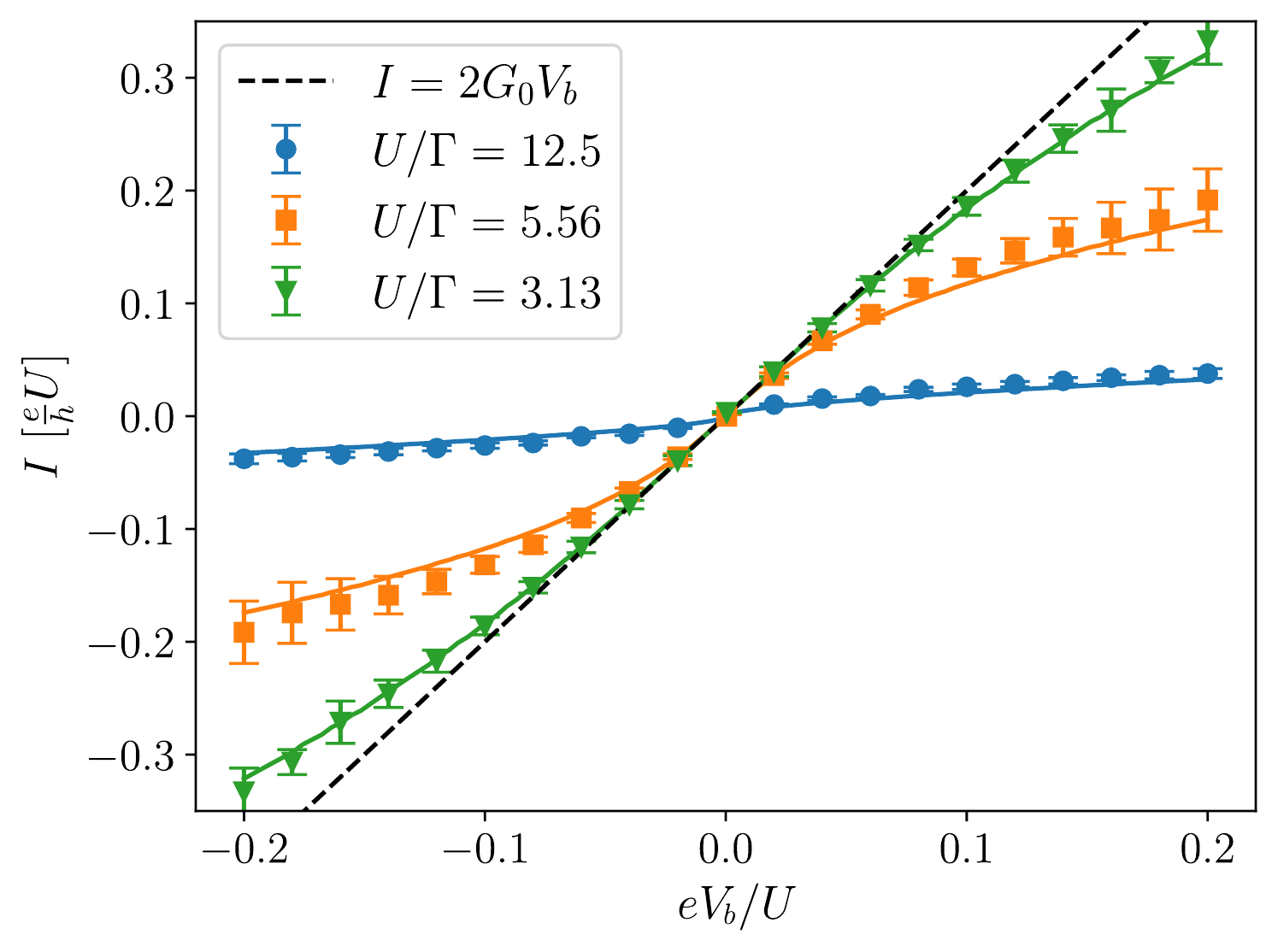}
     \caption{Current vs voltage bias in the symmetric point $V_g=-0.5U$ for three different values of the hybridization strength $\Gamma$ and $U=t_0$. Our data (markers) well match NRG results (solid lines); the temperature of the NRG calculation is $T=10^{-6}U$. The dashed line corresponds to the quantized current $I=2 \frac{e^2}{h} V_b$.}   
\label{fig:curr}
\end{figure}

\begin{figure}
    \centering
    \includegraphics[width=\columnwidth]{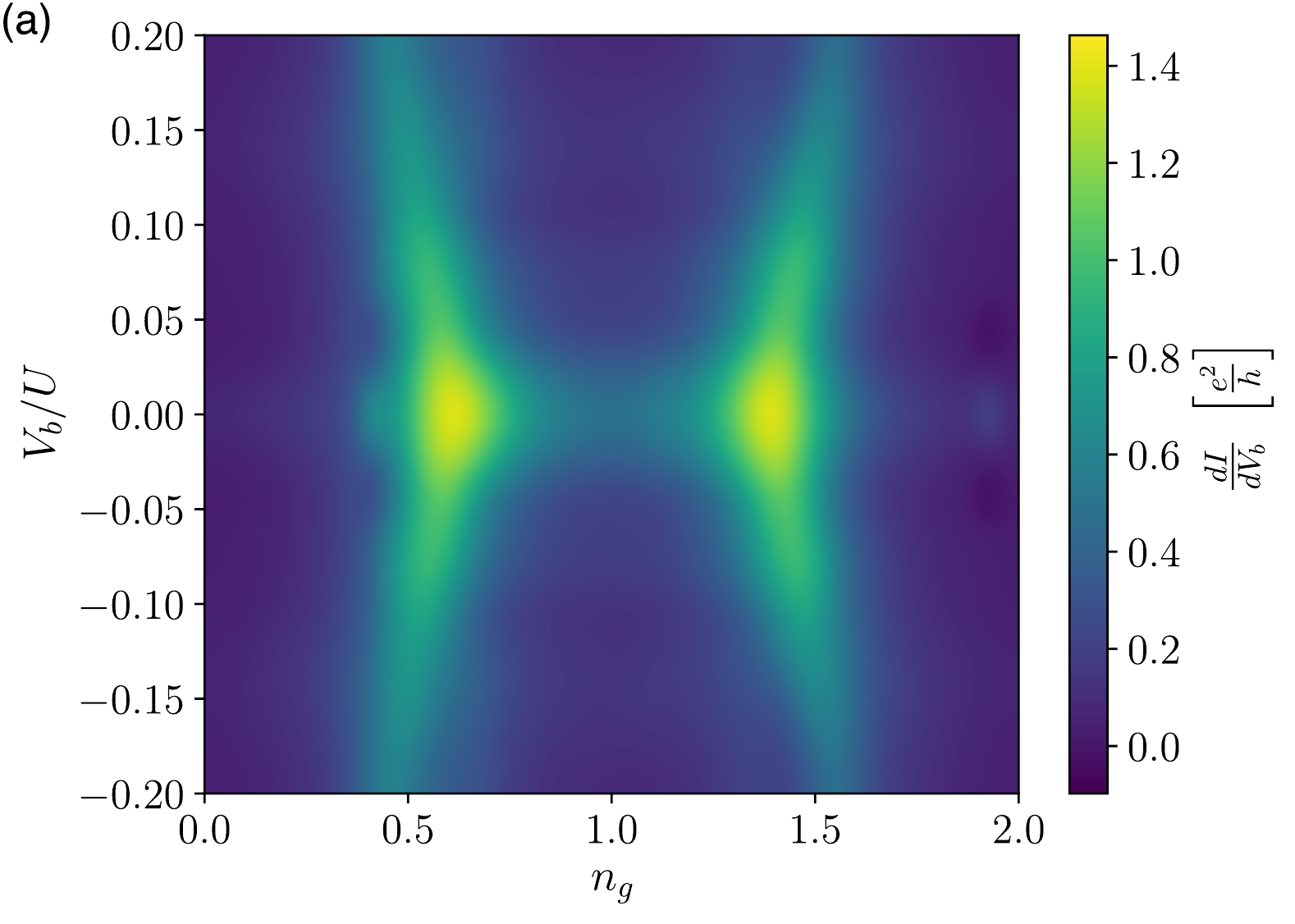}
    \includegraphics[width=\columnwidth]{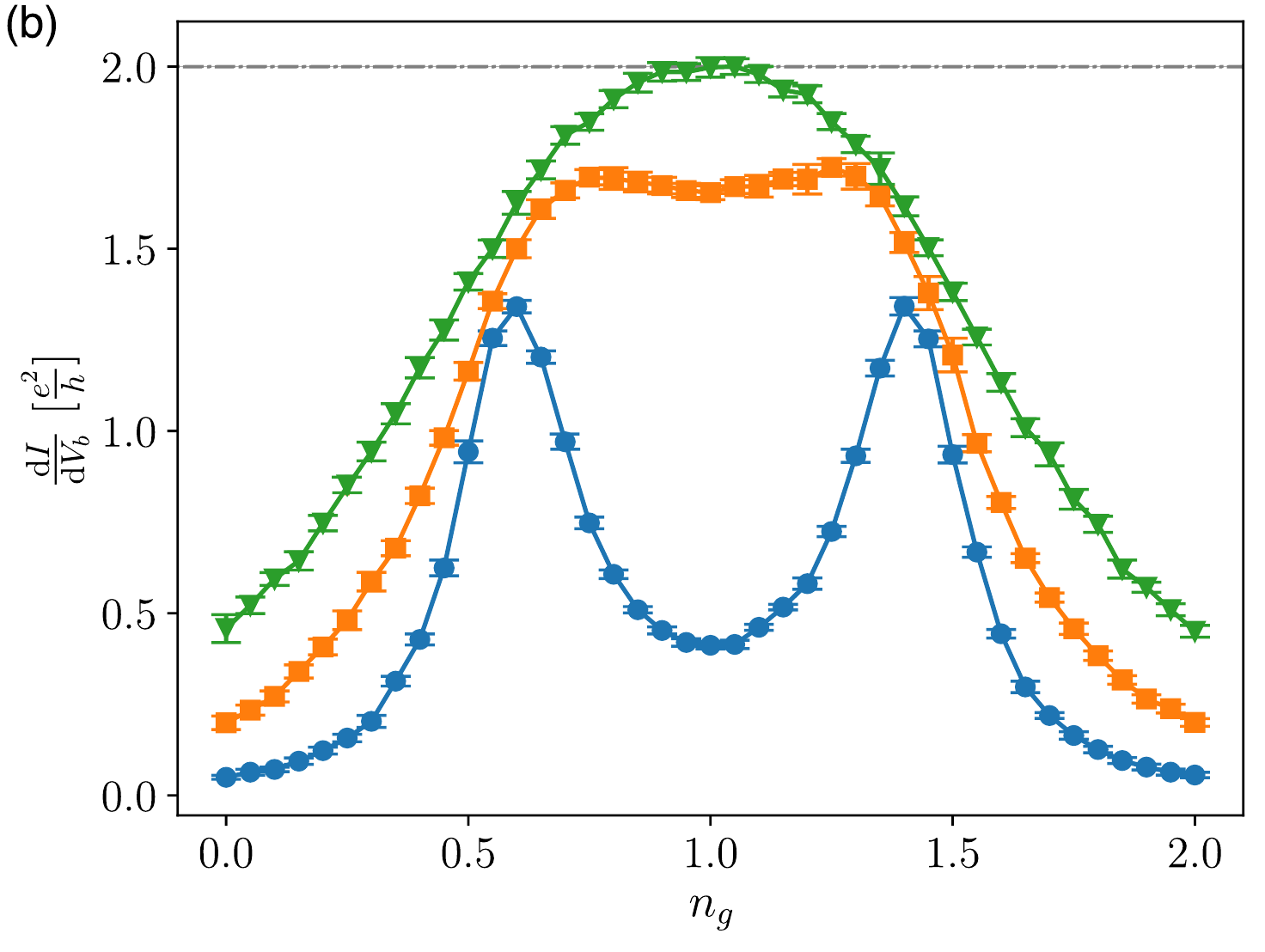}
    \caption{(a) Differential conductance as a function of the induced charge $n_g$ and of the voltage bias $V_b$ between the left and right leads, in the strongly interacting/weak-coupling regime  $U/\Gamma=12.5$. The zero bias peak extending between the two sequential tunneling resonances at $n_g=0.5$ and $n_g=1.5$ signals the onset of the Kondo effect.
    (b) Differential conductance at zero bias as a function of $n_g$, extrapolated with finite bias current obtained at $V_b=0.04U$. The color legend is the same as of Fig.~\ref{fig:curr}.}
    \label{fig:Gdiff}
\end{figure}

The data in Fig. \ref{fig:kondo_relax} are obtained with a zero-bias quench, i.e., with a vanishing bias voltage. When dealing with transport properties, we evolve instead the system with a voltage bias $V_b=\mu_L-\mu_R$ between the two leads in order to observe a quasi-stationary current. 
Here we use the $\mu-$quench protocol: in this scenario the initial state is the correlated ground state of the Hamiltonian $\hat{H}$ (obtained through the DMRG), quenched at ${\sf t}=0$ to a Hamiltonian with a finite voltage bias. 
To simulate the quench dynamics at finite bias, we also need to adjust the decay length of the hopping amplitude in the lead, $\xi$, such that the density of states in the leads is approximately constant in the energy interval between $\mu_L$ and $\mu_R$. 
The convergence in the simulation parameters ($\xi,\ L$, and the TDVP time step discretization) is reached when the current signal displays a plateau in time long enough to reliably extract its expectation value in the NEQSS that develops after the quantum quench. The maximum bond dimension adopted is $\chi=2000$ with a truncation error $O(10^{-8})$.

In Fig.~\ref{fig:curr} we plot the quasi steady current as a function of the voltage bias for different values of the effective tunneling rate $\Gamma$, while keeping fixed the Hubbard interaction $U$, at the particle-hole symmetric point $V_g=0.5U$. 
As we approach the strong-coupling regime $\Gamma\sim U$, the current tends toward a linear response with a quantized differential conductance $\frac{\ud I}{\ud V_b}=2 \frac{e^2}{h}$, i.e. there are two perfectly transmitting channels (dashed black line). 
The Kondo temperature $T_K$ sets the extension of the bias window in which this quantization occurs \cite{Nazarov2000}. In particular, as showed in Fig.~\ref{fig:kondo_relax}(b), the Kondo energy scale drops down exponentially at weak coupling, $U/\Gamma\gg 1$, and can become smaller than the values of voltage bias we can resolve with the chosen lead length $L=100$ and hopping decay length $\xi=30$. This explains the apparent deviation from the quantized conductance at weak coupling $U/\Gamma =12.5$ in Fig~~\ref{fig:curr}.
Our results well match the estimate of the current using the Landauer-B\"uttiker formula and the spectral function of the AIM obtained from NRG calculation~\cite{Zitko_PRB2009,zitko_nrgLjubljana} (solid lines).
Analogous results have been recently obtained in Ref.~\cite{Thoenniss2023} based on an MPS representation, in the temporal domain, of the leads described as Feynman-Vernon influence functionals.
We remark that, away from the strong coupling regime, we can simulate transport for voltages larger than the tunneling rate $\Gamma$. The main limitation comes from the faster entanglement growth when states in a large energy window contribute significantly to transport, which happens when $V_b$ covers a significant fraction of the leads' bandwidth.

Figure~\ref{fig:Gdiff}(a) illustrates the differential conductance in the weak-coupling regime ($U/\Gamma=12.5$) as a function of the bias $V_b$ and the induced charge parameter $n_g$ which is linked to the chemical potential as $V_g = \frac{U}{2}(1-2 n_g)$
and determines the expectation value of the total occupation of the quantum dot. 
 We derive the differential conductance in Fig.~\ref{fig:Gdiff}(a) from the simulation of a $\mu$-quench protocol in which the system is initialized in the ground state of $\Ham$ at half filling, thus for the particle-hole symmetric point ($n_g=1,V_g=-0.5U$).
 At time $\mathsf{t}=0$, both the induced charge $n_g$ and the bias voltage $V_b$ are quenched to their final value [horizontal and vertical axis of Fig. \ref{fig:Gdiff}(a)]. 
 
At $n_g=0.5$ and $n_g=1.5$, we observe two bright zero-bias sequential tunneling resonances, corresponding to the degeneracies between the empty and singly-occupied dot $(n_g=0.5)$ or between the singly and doubly occupied dot $(n_g=1.5)$.
At finite voltage, the conductance peaks are prolonged along the lines $V_b = \pm U(1-2n_g)$ and $V_b = \pm U(3-2n_g)$, following the resonances between each biased lead and the quantum dot.
Between the two charge-degeneracy points, an extended zero bias peak indicates the onset of the Kondo effect, although, for strong interaction and weak coupling, we can not see the quantization of the conductance. 
This limitation originates mainly from the high voltage resolution needed to sample the current at energies below the Kondo temperature, which for $U/\Gamma =12.5$ is of the order $T_K\sim 10{^{-3}} U$, where we expect the quantized linear response. 
To reach such a resolution in $V_b$, we need either a larger system size or a shorter decay length $\xi$. The former makes the simulations computationally more expensive, while the latter induces nonphysical effects in simulations at higher energy, preventing the calculation of the differential conductance in a wide bias range.
As common in nanostructure experiments (see, for instance, Ref. \cite{Jespersen2006}), this zero-bias peak does not extend to $n_g<0.5$ or $n_g>1.5$ where the ground state of the quantum dot becomes, respectively, empty or fully occupied, thus losing the doublet degeneracy necessary for the Kondo effect.

This last feature is particularly evident by looking at the zero-bias conductance as  a function of $n_g$, shown in Fig.~\ref{fig:Gdiff}(b). For all coupling strengths, the conductance is suppressed in the absence of the doublet degeneracy ($n_g<0.5$ or $n_g>1.5$) in the ground state. For $U/\Gamma=3.13$ (green triangles) the quantized peak $\frac{\ud I}{\ud V_b}=2\frac{e^2}{h}$ is reached, as indicated by the dot-dashed grey line. As $\Gamma$ is decreased, the finite bias used to compute the conductance becomes too large to correctly capture the quantization in the linear response regime but the enhancement of the conductance due to the Kondo resonance remains clearly visible for $0.5\le n_g \le 1.5$.
Figure \ref{fig:Gdiff}(b) also exemplifies the expected crossover from a weak-coupling to a strong-coupling regime of the linear conductance extracted from the current at finite DC voltage, as the tuning of $\Gamma$ brings the system deeper into the Kondo regime.

\section{Conclusions}\label{sec:conclusion}
In this work, we applied the tensor network method introduced in Ref.~\cite{Chung_PRB2022} to study the Kondo effect in the Anderson impurity model. In particular, we used an MPS+TDVP approach to study the dynamics of a single-level interacting quantum dot coupled to two fermionic leads after quantum quenches of the Hamiltonian parameters.
We examined both the out-of-equilibrium evolution of the quantum dot magnetization and the electric transport features emerging in a nonequilibrium quasi-steady state after the quench.

The magnetization dynamic allows us to obtain a good estimate of the Kondo temperature as the inverse of its relaxation time when the quantum dot is coupled to unbiased leads. Such estimate is in agreement with renormalization group results \cite{Haldane1978}. In particular, the magnetization decay displays two typical time scales: the effective coupling rate with the leads and the Kondo time scale. The appearance of these two decay regimes for short and intermediate times is reminiscent of the experimental results concerning the evolution of the spin population of impurities in 1D ultracold Yb gases \cite{Riegger2018}. 

Concerning the study of the conductance of the system, relevant for transport measurements in nanostructures, our simulations allow us to study its evolution when a voltage bias is applied between the two leads. By looking at the emergent quasi-steady state, we can reconstruct its Coulomb blockade properties as well as the emergence of a Kondo peak at zero bias. 
The latter appears when the impurity chemical potential fixes its ground state in the degenerate singly-occupied sector and the related differential conductance approaches the quantized value $G=2\frac{e^2}{h}$ in the strong-coupling regime.

We can simulate the system dynamics in a broad parameter range, from a strongly interacting/weak-coupling regime to a strong-coupling one, well beyond the applicability of standard perturbative master-equation approaches.
Moreover, our method is not limited by single-site or small impurities but can be easily extended to multilevel quantum dots or nanowires with long-range Coulomb repulsion.
Despite using Wilson chains to model the leads, our approach does not suffer from problems due to their finite heat capacity~\cite{Rosch2012}. Indeed, our MPS simulations extract transport properties from intermediate-time dynamics where the energy introduced after the quench does not yet translate into a sizeable effective temperature, differently from the eventual asymptotic steady state.

Additionally, our approach can address superconducting systems, opening the path for the study of the out-of-equilibrium dynamics of the topological Kondo effect \cite{Beri2012,Beri2013,Altland2013}, which arises in multiterminal impurities with p-wave superconducting coupling (see Ref. \cite{Wauters_PRB2023}). This kind of system can be easily described by identifying the spin degrees of freedom of the AIM as a label for different spinless leads.

In general, our method can thus be used to examine transport phenomena in hybrid superconducting-semiconducting multiterminal devices with strong Coulomb interactions (see, for instance, Refs. \cite{Kanne2022,Vekris_NanoLett2022}), without being limited to a weak-coupling regime. This offers the possibility of investigating the variety of subgap states ~\cite{Souto_PRB2022} that can appear in these platforms, thus providing important details towards the realization of Majorana - Cooper pair boxes and other building blocks for quantum devices \cite{Oreg2020,Beenakker2020}.

\section*{Acknowledgements}
We thank J. Paaske and V. Baran for fruitful discussions.
M.W., L.M. and M.B. are supported by the Villum Foundation (Research Grant No. 25310). 
This project has received funding from the European Union’s Horizon 2020 research and innovation program under the Marie Skłodowska-Curie Grant Agreement No. 847523 “INTERACTIONS.”
C.-M.C. acknowledges the support by
the Ministry of Science and Technology (MOST) under Grant
No. 111-2112-M-110-006-MY3, and by the Yushan Young
Scholar Program under the Ministry of Education (MOE) in
Taiwan.
This work was supported by Q@TN, the joint lab between
University of Trento, FBK—Fondazione Bruno
Kessler, INFN—National Institute for Nuclear Physics,
and CNR—National Research Council.

\appendix
\section{Spin relaxation: comparison with NRG}\label{app1}
\begin{figure}[h]
    \centering
    \includegraphics[width=\columnwidth]{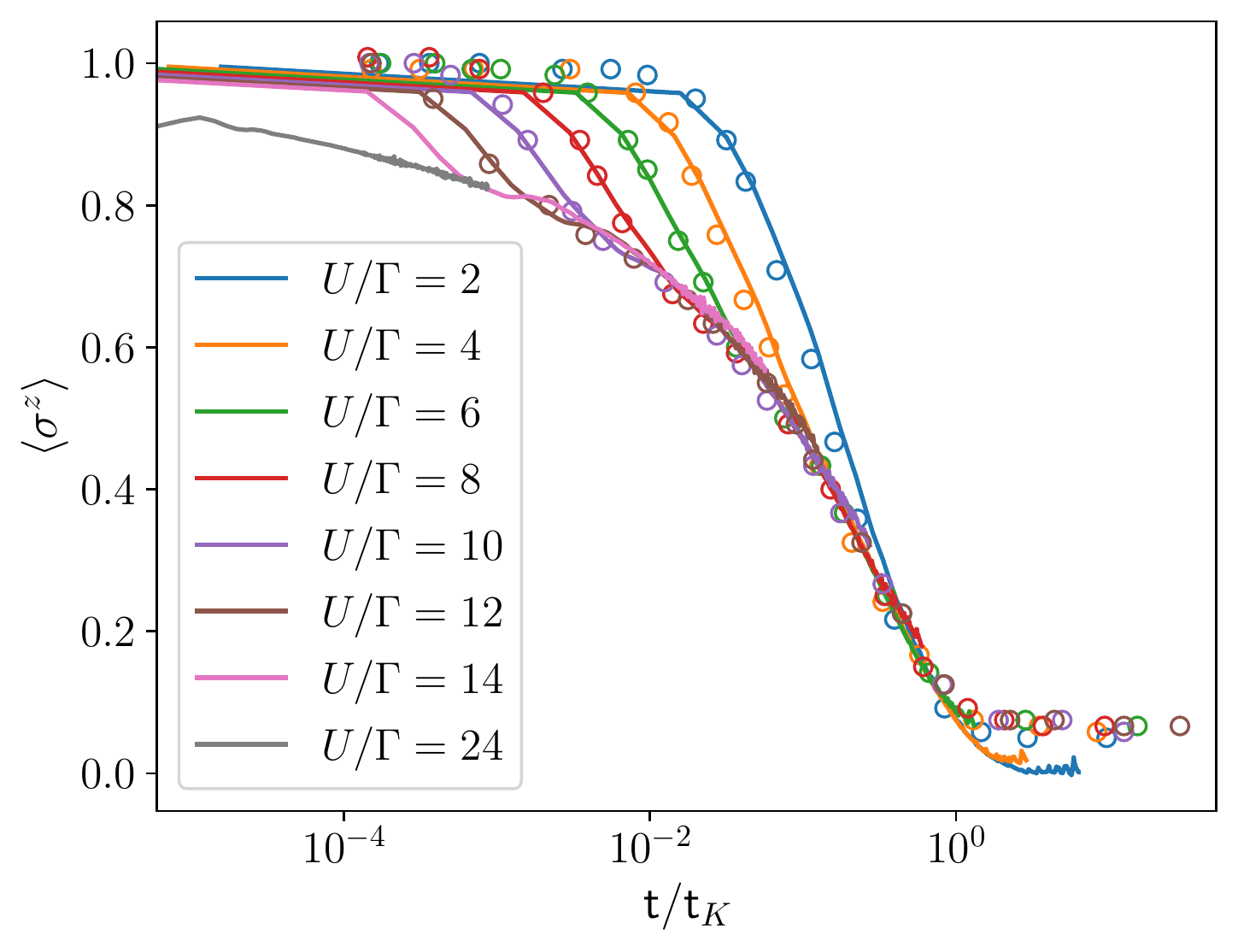}
    \caption{Comparison between the spin relaxation dynamics obtained with the method presented in this paper (solid lines) and the data extracted from Fig.4b of Ref.~\cite{Anders2005} (empty circles). data corresponding to the same ratio $U/\Gamma$ have the same color.}
    \label{fig:NRGcompare}
\end{figure}
As a further benchmark for our code, in this appendix we compare explicitly the results obtained with our MPS method with the seminal time-dependent NRG (TD-NRG) approach introduced in by Anders and Schiller in Ref.~\cite{Anders2005}.
Here we fix the impurity-lead coupling $\Gamma=0.08 t_0$ and change the interaction strength $U$, while the chemical potential is set at the particle-hole symmetric point. The leads have a length $L=100$ and a decay length $\xi=16$. As usual, we use a SVD truncation error of $5\cdot 10^{-8}$ and a maximum bond dimension of 2000. 

In Fig.~\ref{fig:NRGcompare} we show the spin relaxation of the impurity, initially prepared in the $|\uparrow\rangle$ state, in a zero-bias quench.
Curves corresponding to different values of $U/\Gamma$ nicely collapse on top of each other if the time is measured in units of the Kondo time scale ${\sf t}_K$ and we get good agreement with the data extracted from Fig.4b of Ref.~\cite{Anders2005} using the software WebPlotDigitizer~\cite{Rohatgi2022}.  
Notice that we have rescaled their y-axis by a factor of 4 to match our definition of the magnetization.
The main differences between the two approaches occur at small and large times; since we use the same time step $\Delta t =0.5 \hbar/t_0$ in the TDVP algorithm for all values of $U/\Gamma$, we don't have a good resolution at ${\sf t} \ll {\sf t}_K$, in particular when the Kondo time is small. Hence, our simulations do not capture precisely the initial plateau of the magnetization dynamics, as it requires a smaller time step.
At longer times, instead, we observe a relaxation of the magnetization to smaller values, even though the entanglement growth prevents us to reach timescale greater than $\sim 10 {\sf t}_K$.
This discrepancy might be due to the different ways finite-size effects affect TD-NRG and TDVP algorithms.
%

%

\end{document}